%% file: main.tex
\pdfoutput=1
\documentclass[11pt]{article}

\usepackage[]{acl}
\usepackage{times}
\usepackage{latexsym}

\usepackage[T1]{fontenc}
\usepackage[utf8]{inputenc}

\usepackage{microtype}

\usepackage{adjustbox,lipsum}
\usepackage{comment}
\usepackage{enumitem}
\usepackage{tabularx}
\usepackage{booktabs}
\usepackage{array}
\usepackage{float}
\usepackage{amsmath}
\usepackage{hyphenat}
\usepackage{caption}
\usepackage{multirow}
\usepackage{pdfpages}
\usepackage{xpatch}
\usepackage{etoolbox}
\usepackage{color, colortbl, booktabs, makecell}
\definecolor{Gray}{gray}{0.9}
\definecolor{salesassistantblue}{RGB}{11,68,153}
\usepackage[hang,flushmargin]{footmisc}

\usepackage{booktabs}

\DeclareGraphicsExtensions{.pdf,.ai,.jpg,.png}
\setkeys{Gin}{pagebox=artbox}% Alternative (necessary for newer tex versions) for \pdfpagebox 5 in preceding line.
\graphicspath{{./SIGdial23-opinionConv-figures/}}

\newcommand{\Ni}{(1)~}
\newcommand{\Nii}{(2)~}
\newcommand{\Niii}{(3)~}
\newcommand{\Niv}{(4)~}
\newcommand{\Nv}{(5)~}

\RequirePackage{type1cm}% CM-Fonts including Math-Fonts in all sizes.
\RequirePackage{color}
\RequirePackage{soul}
\setstcolor{blue}
\definecolor{lightgray}{rgb}{0.95,0.95,0.95}
\definecolor{lightgreen}{rgb}{0.56,0.93,0.56}
\definecolor{lightblue}{rgb}{0.3,0.3,0.9}
\definecolor{tgray}{rgb}{0.5,0.5,0.5}
\definecolor{red}{rgb}{1.0,0.0,0.0}

\begin{document}
\input{SIGdial23-opinionConv-part0-abstract}
\input{SIGdial23-opinionConv-part1-introduction}
\input{SIGdial23-opinionConv-part2-related-work}

\input{SIGdial23-opinionConv-part3-conversation-corpus}
\input{SIGdial23-opinionConv-part4-human-evaluation}

\input{SIGdial23-opinionConv-part6-conclusion}
\input{SIGdial23-opinionConv-part7-limitations}
\input{SIGdial23-opinionConv-part8-acknowledgements}
\input{SIGdial23-opinionConv-part9-ethics}

\bibliography{custom}
\bibliographystyle{acl_natbib}

%\appendix
%
%\section{Example Appendix}
%\label{sec:appendix}
%
%This is an appendix.

\end{document}

%% file: SIGdial23-opinionConv-part0-abstract.tex
\title{OpinionConv: Conversational Product Search with Grounded Opinions}
%%% NOTES. (use for final version)
%\title{OpinionConv: Conversational Product Search with Grounded Opinions}

\author{Vahid Sadiri Javadi \\
  Conversational AI and Social \\
  Analytics (CAISA) Lab \\
  University of Bonn \\\And
  Martin Potthast \\
  Text Mining and Retrieval \\
  (TEMIR) Group \\
  Leipzig University and ScaDS.AI \\\And
  Lucie Flek \\
  Conversational AI and Social \\
  Analytics (CAISA) Lab \\
  University of Bonn \\}

\maketitle

\begin{abstract}
When searching for products, the opinions of others play an important role in making informed decisions. Subjective experiences about a product can be a valuable source of information. This is also true in sales conversations, where a customer and a sales assistant exchange facts and opinions about products. However, training an AI for such conversations is complicated by the fact that language models do not possess authentic opinions for their lack of real-world experience. We address this problem by leveraging product reviews as a rich source of product opinions to ground conversational AI in true subjective narratives. With OpinionConv, we develop the first conversational AI for simulating sales conversations. To validate the generated conversations, we conduct several user studies showing that the generated opinions are perceived as realistic. Our assessors also confirm the importance of opinions as an informative basis for decision making.
\end{abstract}

%% file: SIGdial23-opinionConv-part1-introduction.tex
\section{Introduction}

\begin{figure*}[!ht]
\centering
\includegraphics[width=\textwidth]{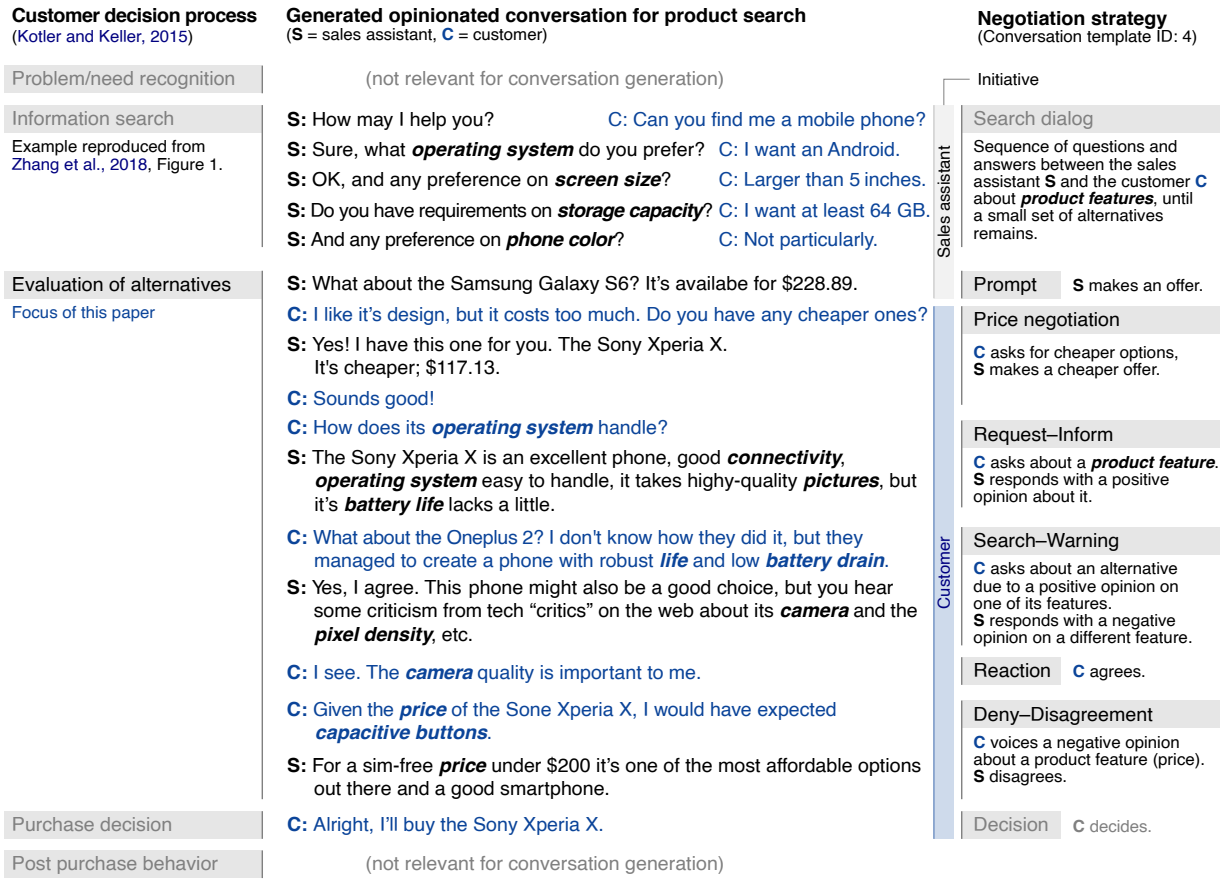}
\caption{A grounded opinionated conversation generated by OpinionConv based on Conversation Template~4.}
\label{sales-negotiation-scheme}
\end{figure*}

In order to elucidate the mechanics of conversational product search, \citet{kotler:2015} delineated a five-stage process that encapsulates customer decision making (see Figure~\ref{sales-negotiation-scheme}, left). This process suggests that the customer:
\Ni
recognizes a problem or need;
\Nii
searches for information about potential products or services that could resolve the problem or fulfill the need, filtering them until a manageable set of alternatives remains;
\Niii
evaluates and compares these alternatives against each other with regard to personal preferences and third party opinions to inform their decision making;
\Niv
proceeds to make a purchase decision predicated upon this informed evaluation; and finally,
\Nv
exhibits post-decision behaviors that reflect their satisfaction, which completes the process.

Typically, in-store shopping predominantly engages with the second and third stages of this customer decision process. Both the activities of reducing the number of alternatives and evaluating their merits and demerits are conducted in conversations between customers and sales assistants. The absence of such interactions in online environments is perceived as a deficiency in customer service especially with respect to the third stage \cite{exalto:2018}. Customers derive post-purchase satisfaction from personal exchanges, relating to others experience, and having the opportunity to ask questions \cite{papenmeier:2022}. The considerable number of online product reviews available are not a substitute for everyone, since many customers lack the patience to examine many of them, leading to post-purchase dissatisfaction and product returns. Conversational~AI has been suggested as a solution \cite{gnewuch:2017}, with the goal of mimicking the conversational strategies of sales assistants \cite{papenmeier:2022}. But despite its importance, previous research on conversational product search almost entirely neglects the third stage, or rather its opinionated aspects (Section~\ref{related-work}).

Recent advances in large-scale conversational language models, spearheaded by OpenAI's ChatGPT, are driving a paradigm shift in the development of conversational technologies. Nonetheless, when it comes to expressing opinions pertaining to real-world events or entities, these language models lack the necessary grounding in tangible reality. For an individual to formulate an opinion on a particular subject matter, they require exposure to the subject to relate the new experience to past ones, and importantly, an emotional perception. A language model is only capable of generating what might be termed as a ``statistical average'' of third-party opinions, if they have been part of its training data. In the context of product search, such opinions would be deemed unauthentic as they are not based on real-world experiences or substantiated knowledge. This lack of authenticity poses challenges to the effective utilization of these models when (personal) opinions play an important role.

In this paper, we focus on the third stage of the customer decision process, for which we contribute the first approach to generate grounded opinionated statements (Section~\ref{opinion-generation-for-products}). We conceive and operationalize the generation of grounded opinions by positing that a grounded opinion about a product is an opinion which has been verifiably expressed by a minimum of one individual in a product review that specifically discusses the product under scrutiny. Our approach, OpinionConv, combines a product-specific index of reviews for a cohort of products of the same kind with a mechanism to generate realistic opinionated conversational exchanges. While carefully tuned, our approach must still be considered an early prototype. Consequently, before asking real customers to use it, its fundamental capabilities must first be established. We therefore simulate in-store dialogues between a customer and a sales assistant, where both parties incorporate grounded opinions. These conversations are then systematically evaluated in an experimental setup that ascertains the perception of human readers regarding the realism of these dialogs (Section~\ref{evaluation}).%
\footnote{Code and data: \url{https://github.com/caisa-lab/OpinionConv}}

%% file: SIGdial23-opinionConv-part2-related-work.tex
\section{Related Work}
\label{related-work}

Three lines of research are related to ours: opinionated question answering, conversational product search, and review-based conversation generation.

\subsection{Opinionated Question Answering}

While factoid Question Answering~(QA) systems have a long tradition and some even outperform humans, non-factoid questions, such as opinions, explanations, or descriptions, are still an open problem \cite{cortes2021systematic}. \citet{cardie2003combining} employed opinion summarization to help multi-perspective QA~systems identify the opinionated answer to a given question. \citet{yu2003towards} separated opinions from facts and summarized them as answers. The linguistic features of opinion questions have also been studied \cite{pustejovsky2005introduction, stoyanov2005multi}. \citet{kim2005identifying} identified opinion leaders, which are a key component in retrieving the correct answers to opinion questions. \citet{ashok:2020} introduced a clustering approach to answer questions about products by accessing product reviews. \citet{rozen:2021} examined the task of answering subjective and opinion questions when no (or few) reviews exist. \citet{jiang2010framework} proposed an opinion-based QA~framework that uses manual question--answer opinion patterns.

Closer to our work, \citet{moghaddam2011aqa} address the task of answering opinion questions about products by retrieving authors' sentiment-based opinions about a given target from online reviews. \citet{mcauley2016addressing} address subjective queries using relevance ranking, and \citet{wan2016modeling} extends this work by considering questions that have multiple divergent answers, incorporating aspects of personalization and ambiguity. AmazonQA \cite{gupta2019amazonqa} is one of the largest review-based QA~datasets. Its authors show that it can be used to learn relevance in the sense that relevant opinions are those for which an accurate predictor can be trained to select the correct answer to a question as a function of opinion. SubjQA \cite{bjerva2020subjqa} includes subjective comments on product reviews.

\begin{figure*}
\includegraphics[width=\textwidth]{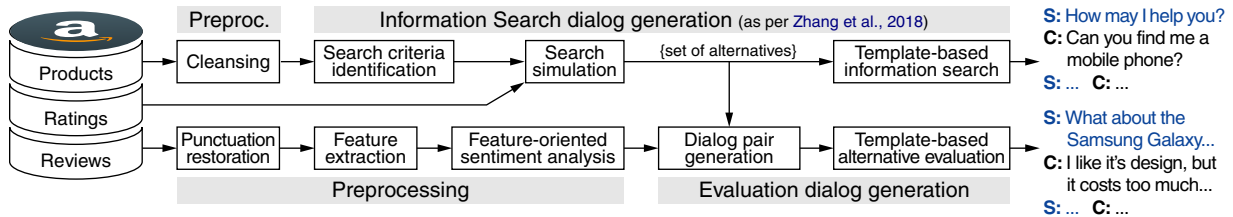}
\caption{High-level overview of our approach in OpinionConv for generating opinionated multi-turn conversations.}
\label{sales-conversation-generation-pipeline}
\end{figure*}

\subsection{Conversational Product Search}

Information is often gathered through conversations with a series of questions and answers. Conversational Question Answering~(CQA) systems engage in such multi-turn conversations to satisfy a user's information need \cite{zaib2021conversational}. Despite the attention this task has received in e-commerce \cite{ricci2011introduction, bi2019conversational, zhang2018towards}, building a successful conversational product search system for online shopping still suffers from the lack of realistic dialog datasets for model training \cite{xiao2021end}.

\subsection{Review-based conversation generation}

Recently, multi-turn~QA has grown more prominent \cite{cambazoglu:2021}. Product reviews are one of the sources of information that are being used for conversational product search. \citet{penha:2022} generate review-based explanations for voice-driven product search. \citet{zhang2018towards} builds a dataset to answer conversational questions, as illustrated in Figure~\ref{sales-negotiation-scheme}~(Stage~2 ``Information Search''). They extract feature--value pairs from reviews and convert each review into a conversation based on the mentioned pairs, but omit opinionated statements. \citet{xu2019review} explores the possibility of turning reviews into a knowledge source to answer questions. Feature-related, non-opinionated statements in reviews are flagged and appropriate questions are formulated.

%% file: SIGdial23-opinionConv-part3-conversation-corpus.tex
\section{Grounded Product Opinion Generation}
\label{opinion-generation-for-products}

This section introduces the OpinionConv construction pipeline to generate grounded opinionated conversations for product search based on product reviews. Figure~\ref{sales-conversation-generation-pipeline} gives an overview of the pipeline's individual steps, grouped into preprocessing, information search dialog generation (Stage~2 of the customer decision process, which we reproduce from \citet{zhang2018towards}), and evaluation dialog generation (Stage~3, our focus).

\subsection{Data Source and Preprocessing}
\label{data source and preprocessing}

As a basis for grounded opinions, we utilize a crawl of Amazon product data including their reviews created by \citet{ni2019justifying}.%
\footnote{\url{https://jmcauley.ucsd.edu/data/amazon/}}
The metadata enclosed includes product descriptions, multi-level product categories, and product information. For our proof-of-concept, we focus on one of its 24~product categories, \emph{Cell Phones and Accessories}. As a first cleansing step, we reviewed the product data and added missing product details. We found the reviews to be of varying writing quality, especially with respect to basic syntax conventions, like the use of punctuation. We employed the model of \citet{alam2020punctuation} to restore the punctuation, which enabled a more reliable sentence extraction and thus benefited the subsequently applied models, which were largely trained on ``cleaner'' data.

To extract the product features discussed in the reviews, we use the extraction model of \citet{karimi:2021}. It is based on a hierarchical aggregation approach and was trained on the laptop review dataset of SemEval~2014 \cite{pontiki:2014}, performing best at that time. Given a review sentence, the model extracts feature terms on which an opinion has been expressed. On sentences containing such opinion statements, we then applied the sentiment analysis model of \citet{zeng:2019}, which is based on self-attention to capture local context and global context features to determine the polarity score of the opinion.

\subsection{Information Search Dialog Generation}

To generate the information search dialog of the customer decision process (see Figure~\ref{sales-negotiation-scheme}), we reproduce the approach of \citet{zhang2018towards}. Reproducing the original dialogs turned out to be straightforward, and we verified our success by direct comparison to the data supplied with the original paper. The dialogs are structured as follows: The sales assistant asks for preferences on product features, and the customer answers, narrowing down the set of alternatives. The resulting set of alternatives is fed to the next stage of the customer decision process, the evaluation of alternatives.

\subsection{Evaluation Dialog Generation}

The generation of an opinion-based evaluation of alternatives dialog is divided into two steps, the generation of pairs of talking points based on reviews of the alternative products, and their combination into a multi-turn conversation as exemplified in Figure~\ref{sales-negotiation-scheme}. For lack of public corpora of in-store conversations, we resort to a template-based approach. The templates are derived from common conversational negotiation strategies from the literature.

\paragraph{Dialog Turn Pair Generation}
In each turn of an evaluation dialog the customer and the sales assistant discuss the relative value of product features, as well as their benefits and shortcomings compared to alternative products. Sales assistants, by training and/or experience, are usually well-equipped to provide customers with satisfying answers to their questions as well as to respond to opinions that customers express throughout the conversation. The most salient open question in this respect is: How should a sales assistant react to a customer's opinion in the context of a negotiation? 

\begin{figure}
\centering
\includegraphics{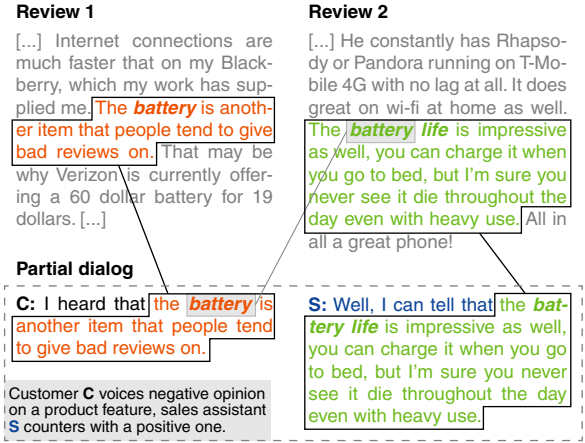}
\caption{Example of a basic opinionated dialog pair generation step: Given a product feature like ``battery'', opinionated statements are extracted from reviews of a given product to form part of a dialog between Customer~C and Sales Assistant~S.}
\label{dialog-pair-extraction}
\end{figure}

\input{table-negotiation-tactics-for-dialog-pairs}

Negotiations combine features of claiming and creating value. Each requires unique strategies and tactics for a negotiator to effectively achieve their objectives while creating the greatest value possible for all parties \cite{thompson:1990}. We take inspiration from three negotiation tactics \cite{dwyer:1987, suavescu2019some}:
(1)~\emph{Distributive negotiation}:
This is a competitive win--lose situation. Any value claimed by one party is at the expense of the other.
(2)~\emph{Integrative negotiation}:
The parties create or generate value during the negotiation, and both parties may achieve mutual gains beyond what they would achieve independently, a win--win scenario.
(3)~\emph{Compatible negotiation}:
The parties desire the exact same outcome, so that there is no need for any trade-off.

For instance, as illustrated in Figure~\ref{dialog-pair-extraction}, where Customer~C's remark about a product feature~(left) is countered by an opinionated counterargument from Sales assistant~S~(right), we use the Deny--Disagreement tactic, where the customer expresses a negative opinion on a feature of a product in question, whereas the sales assistant disagrees and counters with a positive opinion on the same feature. This tactic may either correspond to a win--lose or a win--win situation, dependent on whose opinion applies more to the customer: If the customer is correct, they lose against the sales assistant, since the product is not switched. If the sales assistant is correct, they both win, since the customer still gets what they wanted, and the sales assistant may still get to sell the product in question.

A key constraint that we enforce by generating grounded opinions (i.e., opinions rooted in a real product reviews as exemplified in Figure~\ref{dialog-pair-extraction}) is that neither the customer nor the sales assistant can ``lie'' to each other, as their opinions are backed by a real person's opinion about the product and its feature. Thereby the dialog turns are more realistic, despite both parties being simulated. Moreover, our dialog turns enforce a conversational concept flow between \cite{li:2023}, as the product features and their attributes as key concepts are connected.

In Table~\ref{table-negotiation-tactics-for-dialog-pairs}, we list the templates  for dialog pairs according to different negotiation tactics derived from the literature; seven patterns are devised, one question--answer pair, and six opinion--opinion pairs. The customer's utterance consists of a feature-specific opinion with either positive or negative opinion for a certain product and one of its features, extracted from one of its reviews. The sales assistant's utterance is a response that expresses either a positive or a negative opinion, not necessarily to the same product or feature.

\input{table-conversation-template-example}

As can be seen, depending on the type of dialog pair, different negotiation tactis may apply. The sales assistant is allowed to switch the polarity, feature under discussion, and product under negotiation. The mapping of a dialog pair to the negotiation tactics thus depends on the factuality of either opinion expressed by the customer or sales assistant, in case a product switch on the price changes (the price of the new product may be lower, similar, or higher), but also on whether the customer gets what they want. For instance, a switch to a pricier product is certainly worthwhile for the sales assistant, as long as the customer ends up with a desired feature. However, given the necessity of interpreting each generated dialog with respect to its factuality, a mapping between dialog pair and negotiation tactic must be decided on a case-by-case basis, which is beyond the scope of this paper.

A constraint in cases where the product is being changed includes that the sales assistant is allowed to use only two types of products: 
\Ni
Retrieved products: The products from the set of alternatives retrieved in the information search dialog at the outset of a conversation based on customer preferences;
\Nii
also viewed products: The products listed in the metadata that have been viewed by other customers. Customers thus can go beyond their original preferences and the set of alternatives.

\paragraph{Template-based Alternative Evaluation}
The last step of our pipeline generates conversations composed of multiple dialog pairs, based on a negotiation strategy. Considering the diversity of real dialogs and the fact that a coherent conversation should have a smooth transition between turns \cite{li:2023}, we define a diverse set of conversation templates inspired by past studies on negotiation in behavioral economics \cite{pruitt:1981, fisher:1981, thompson:2010}, including both high-level (e.g., insisting on your position: Disagreement) and low-level (e.g., focus on interests: Reaction) dialog acts.

We follow \citet{zhou:2019} and devise 14~conversation templates with different combinations of the generated question--answer and opinion--opinion pairs. Table~\ref{table-conversation-template-example} exemplifies one of them. We adapt the ``CraigslistBargain'' setting of \citet{he:2018}, where a buyer and a seller negotiate the price of a product. But unlike in their work, the sales assistant and the customer negotiate not only the price but primarily the relative merits of product features, whereas price may only be one of them.

%% file: table-negotiation-tactics-for-dialog-pairs.tex
\begin{table*}[t]
\centering
\small
\renewcommand{\tabcolsep}{12pt}
\caption{Negotiation tactics used in dialog pair templates (P=product, F=feature).}
\label{table-negotiation-tactics-for-dialog-pairs}
\begin{tabular}[t]{@{}lp{11.5cm}@{}}
\toprule
\textbf{Dialog pair template} & \textbf{Description of negotiation tactic} \\
\midrule
\begin{tabular}[t]{@{}l@{}} 
\textbf{Request--Inform} \\ 
Question: P-1, F-A, neutral \\ 
Answer: P-1, F-A, positive
\end{tabular} & 
Customer asks about the sales assistant's view on a feature of a product. Sales assistant expresses positive view on it. \\
\midrule
\begin{tabular}[t]{@{}l@{}} 
\textbf{Deny--Disagreement} \\ 
Opinion: P-1, F-A, negative \\ 
Opinion: P-1, F-A, positive
\end{tabular} &
Customer expresses negative opinion on a feature of a product. Sales assistant disagrees and expresses positive opinion on it.\\
\midrule
\begin{tabular}[t]{@{}l@{}} 
\textbf{Deny--Switch Product} \\ 
Opinion: P-1, F-A, negative \\ 
Opinion: P-2, F-A, positive
\end{tabular} &
Customer expresses negative opinion on a feature of a product. Sales assistant switches the product and expresses positive opinion on the same feature wrt.\ new product. \\
\midrule
\begin{tabular}[t]{@{}l@{}} 
\textbf{Deny--Switch Feature} \\ 
Opinion: P-1, F-A, negative \\ 
Opinion: P-1, F-B, positive
\end{tabular} &
Customer expresses negative opinion on a feature of a product. Sales assistant disagrees and expresses positive opinion on a different feature of the same product. \\
\midrule
\begin{tabular}[t]{@{}l@{}} 
\textbf{Search--Agreement} \\ 
Opinion: P-1, F-A, positive \\ 
Opinion: P-1, F-A, positive
\end{tabular} &
Customer expresses positive opinion on a feature of a product. Sales assistant agrees and expresses another positive opinion it. \\
\midrule
\begin{tabular}[t]{@{}l@{}} 
\textbf{Search--Switch Feature} \\ 
Opinion: P-1, F-A, positive \\ 
Opinion: P-1, F-B, positive
\end{tabular} &
Customer expresses positive opinion on a feature of a product. Sales assistant agrees and expresses positive opinion on different features of the same product. \\
\midrule
\begin{tabular}[t]{@{}l@{}} 
\textbf{Search--Warning} \\ 
Opinion: P-1, F-A, positive \\ 
Opinion: P-1, F-B, negative
\end{tabular} &
Customer expresses positive opinion on a feature of a product. Sales assistant warns the user and expresses negative opinion on different features of the same product. \\
\bottomrule
\end{tabular}
\end{table*}

%% file: table-conversation-template-example.tex
\begin{table*}
\centering
\small
\renewcommand{\tabcolsep}{4pt}
\caption{Example of the combination of dialog pairs in a conversation template.}
\label{table-conversation-template-example}
\begin{tabular}{@{}p{2cm}p{2.5cm}p{2.6cm}p{8cm}@{}}
\toprule
\bfseries Pair & \bfseries Principle & \bfseries Action & \bfseries Example \\
\midrule
\raggedright Deny--Switch Product 
& \raggedright Insist on position 
& \raggedright Express negative sentiment 
& {\bfseries B:} What I know about its battery is that the battery keeps draining because the phone is constantly looking for network signal. \\
& \raggedright Invent options for mutual gain 
& \raggedright Recommend a new product 
& \textcolor{salesassistantblue}{{\bfseries S:} If the battery is important for you, we can offer this product: Axon~7 is the same price as OnePlus 3, but it has slightly bigger battery.} \\
\midrule
\raggedright Request--Inform 
& \raggedright Focus on interests 
& \raggedright Look for more information 
& {\bfseries B:} What do you think about its speakers? \\
& \raggedright Build trust 
& \raggedright Express positive sentiment 
& \textcolor{salesassistantblue}{{\bfseries S:} It has dual front-facing speakers with good quality.} \\
\midrule
\raggedright Search--Agreement 
& \raggedright Focus on interests 
& \raggedright Search for alternatives 
& {\bfseries B:} I heard about this phone: Galaxy S4 that has a super-fast processor and  a good battery life. \\
& \raggedright Build trust 
& \raggedright Confirm consumer's preference 
& \textcolor{salesassistantblue}{{\bfseries S:} Yes, that's true. This phone is also a good choice with the one premium hardware, great software and a reasonable price.} \\
\bottomrule
\end{tabular}
\end{table*}

%% file: SIGdial23-opinionConv-part4-human-evaluation.tex
\section{Evaluation}
\label{evaluation}

\input{table-corpus-statistics}

\enlargethispage{-\baselineskip}
A volunteer who is asked to pose as a customer in a laboratory user study, and who has no real intention of investing a fairly large amount of his or her own money in the purchase of a product, does not usually have the same information needs as a real customer. At the same time, we consider it unethical to confront real customers with an early prototype of a conversational sales assistant. Before a practical assistant can be developed, the basic means of generating informed opinions must first be established.

In our evaluation, we therefore decided to simulate full conversations between a hypothetical customer and a hypothetical sales assistant as described in the previous section. We then designed two user studies in which we specifically investigated whether human subjects consider these conversations realistic. 

Study~1 investigates whether subjective narratives in conversational product search are considered important compared to purely factual exchanges. Study~2 investiages individuals’ perceptions of the quality and realism of the conversations generated by OpinionConv. We conducted the two studies by recruiting volunteers living in the~US or~UK on Prolific.%
\footnote{\url{https://www.prolific.co}}
Table~\ref{table-corpus-statistics} shows the total sample size and key demographic information for both studies. As can be seen, we had fewer male participants than females and more than 60\% of the participants are between 25~and 45~years old. Key to both our study design is that participants initially believed that the conversations are genuine transcripts of real sales negotiations recorded in a store, instead of generated ones. At the end of the questionnaire, it was revealed that they are not.

\subsection{Study~1: Importance of Product Opinions}

The first study started with the following instruction: ``Below is an automatically generated transcript of a sales conversation. We show two variants: Variant~1 is focused on the customer’s preferences and requirements. Variant~2 starts similarly, but then continues with an opinionated discussion.'' After reading both variants of the same dialog participants were asked ``Which of the two variants would you as a customer hold with the sales assistant while searching for a smartphone?'' The survey concluded by asking participants an open-ended question to explain their judgment for the previous question, which also allowed to ascertain that they had actually read the conversations.

\begin{figure*}
\centering
\includegraphics[width=\textwidth]{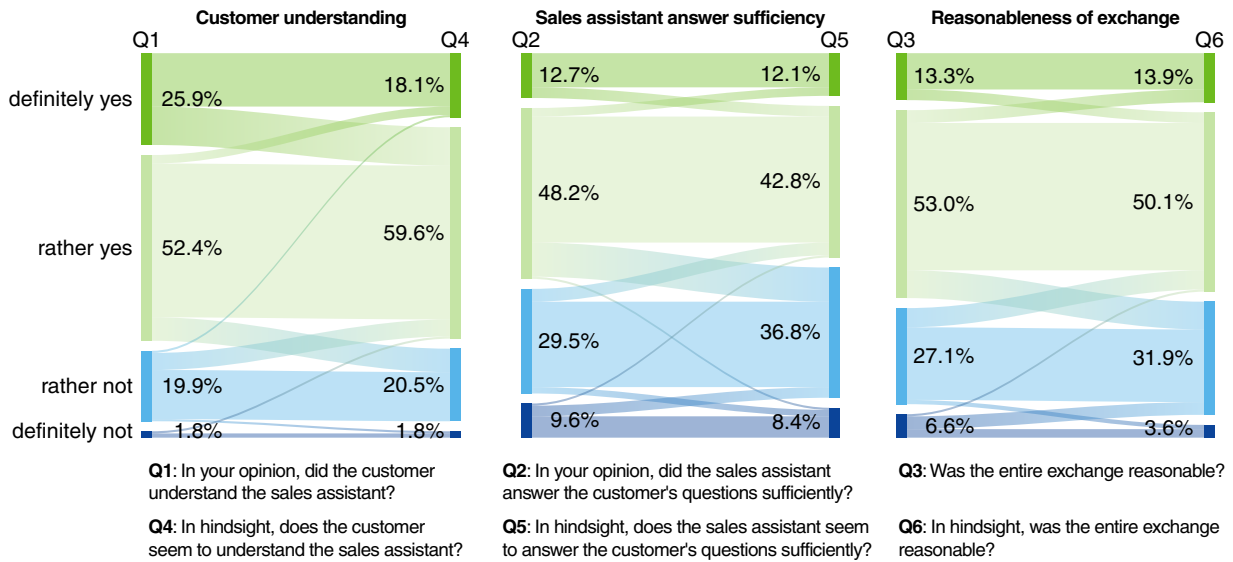}
\caption{Evaluation results for Study~2; Questions~Q1, Q2, and~Q3 are asked in the first part of the questionnaire (before disclosing the conversations are generated), Q4, Q5, and~Q6 are asked in the second part (after disclosure).}
\label{plot-conversation-assessment}
\end{figure*}

As a result, we find that~83\% of the 100~participants of Study~1 prefer Variant~2 over the Variant~1, which confirms that they tend to prefer opinionated conversations when searching and evaluating a product rather than exclusively factual ones.

\subsection{Study~2: Perceptions of Dialog Realism}

For this study, the questionnaire consisted of two separate parts. In the first part, we again let participants believe they are reading a transcript of a real conversation by instructing them as follows: ``Suppose you are in an electronics store. While browsing, you happen to overhear part of a conversation between a customer and a sales assistant. Both exchange opinions about the features of one or more products.'' They are then asked to answer three questions using the 4-point Likert scale: (1)~\textit{Definitely yes}, (2)~\textit{Rather yes}, (3)~\textit{Rather not} and (4)~\textit{Definitely not}. As depicted in Figure~\ref{plot-conversation-assessment}, bottom, we ask questions for the following goals: \textit{Customer understanding}, \textit{Sales assistant answer sufficiency} and \textit{Reasonableness of exchange}. To investigate whether participants' perceptions change significantly after they learned that the conversation was generated, at the beginning of the second part, we reveal the truth and declare that the conversation they just read, was not a real but an automatically generated one. After the disclosure, they were asked answer Questions~4 to~6 using the same Likert-scaled responses as shown in Figure~\ref{plot-conversation-assessment} in order to observe any changes of opinion. For each of our 14~conversation templates, we generated ten examples, and for each example, three participants were asked to answer the questions, a total of 140~questionnaires answered by 420~participants.

Figure~\ref{plot-conversation-assessment} shows the distribution of participants' responses to each question, outlining the alterations in perception after revealing the automated generation of conversations. Both user (Q1~\&~Q4) and agent (Q2~\&~Q5) utterances, as well as the overarching dialog (Q3~\&~Q6), were subjected to quality evaluation. The data reveals that prior to revealing the truth, over 66\% of evaluators deemed the conversation reasonable (both ``yes'' answers combined), marginally reducing to 64\% post-revelation (Q3~\&~Q6). Regarding participants' assessment of the customer's understanding of the sales assistant (Q1~\&~Q4), over 78\% affirmed it, which reduces to 77\% after the disclosure, albeit almost half of participants switched form \textit{definitely yes} to \textit{rather yes}. Regarding the evaluation of the sales assistant's response quality to customer inquiries (Q2~\&~Q5), over 60\% of participants agreed that the responses were sufficient, while the disclosure incited a reduction in both \textit{definitely yes} and \textit{rather yes} responses to over 54\%. Altogether, the responses indicate a generally positive reception of conversations generated by OpinionConv, with variation in assessment among different conversation templates. While a rough two thirds of participants agree with this outcome, more than half of participants improve their rating from \textit{definitely not} to \textit{rather not} considering overall reasonableness.

\subsection{Participants' Comments}
\label{participants comments}

In both studies, participants were asked to explain their judgment in about two sentences, in case of Study~2 once in each part of the questionnaire. With respect to assessing the conversation realism, we mostly observe positive comments. For instance, before disclosing the nature of the conversation one participants commented ``The customer was recommended the phone by a friend, and the sales assistant was able to give further information on the phone. Likewise the sales assistant was able to inform the customer on a drawback associated with the phone.'', and after ``The conversation appeared real as there was flow - i.e. the sales assistant was able to connect with what the customer said and elaborate upon it. Likewise the sales assistant was able to pick up on key details associated with the phone like the camera and OS.''

%%% ADDITION.
%%% Before disclosure.
%``I think the exchange was reasonable and I think the sales assistant did their job properly. They gave their personal opinion, told the truth, and answered the customer's questions to the best of their ability.''
%%%
%%% After disclosure.
%``Even now knowing that the conversation wasn't real, I think the exchange seems reasonable. The customer asked questions about the product and the sales assistant answered them properly (they could have elaborated on the design question though).''

However, we also observe three key concerns raised:
\Ni
Some features are of no interest to be discussed, e.g., ``Why would the person asks the sales assistant about colors? That seems out of the ordinary.''
\Nii
Some participants judge the conversations based on their personal experience with real sales assistants, e.g., ``As always in marketing strategies, he [the sales assistant] was just trying to sell a phone not what he [the customer] wanted.''
\Niii
A stronger argumentation is expected by some, e.g., ``While the sales assistant did respond in a way that does answer the customer's questions, their responses are not so direct and detailed as to be helpful towards the customer. For example, for the question about the screen, stating that it's `bright and good quality' would not be convincing enough for me to want to buy the product.''

Reading the participants' comments and observing the results of crowd-sourced qualitative evaluations have suggested several new research directions for future work relating to common sense product knowledge and argument generation.

%% file: table-corpus-statistics.tex
\begin{table}[t]%
\centering%
\small%
\renewcommand{\tabcolsep}{10pt}%
\caption{Demographics of study participants.}
\label{table-corpus-statistics}
\begin{tabular}[t]{@{}llrr@{}}
\toprule
  \textbf{Measure} & \textbf{Characteristics} & \textbf{Study 1} & \textbf{Study 2} \\
                   &                          & (N=100)          & (N=420)          \\
\midrule
  Gender           & Males                    & 41.0\%           & 31.0\%           \\
                   & Females                  & 58.0\%           & 69.0\%           \\
                   & Non-binary               & 1.0\%            & 0.0\%            \\
\hline
  Age              & 25 to 34 years           & 35.0\%           & 38.0\%           \\
                   & 35 to 44 years           & 28.0\%           & 30.1\%           \\
                   & 18 to 24 years           & 21.0\%           & 15.7\%           \\
                   & 55 to 64 years           & 6.0\%            & 13.3\%           \\
                   & 45 to 54 years           & 5.0\%            & 1.8\%            \\
                   & 65 years or older        & 5.0\%            & 1.2\%            \\
\bottomrule
\end{tabular}%
\end{table}

%% file: SIGdial23-opinionConv-part6-conclusion.tex
\section{Conclusion}

We introduce the OpinionConv, a new conversation generation pipeline that generates opinionated multi-turn conversations for product search. OpinionConv was mainly designed to incorporate subjective narratives into conversational product search. The pipeline presented in this work can be easily extended to different domains. Recent progress in conversational systems, such as ChatGPT and YouChat, have shown tremendous improvements in natural language dialog between humans and conversational agents. However, when it comes to holding an opinionated conversation, specifically in product search, they are still limited for lack of grounding in real-world experience about products. This motivated the design of a pipeline to control both the dialog coherence and the information to be mentioned in the utterances. However, it should be mentioned that the trade-off between a coherent conversation and a more diverse conversation needs to be further studied. In order to validate the quality of the conversations generated by OpinionConv, we conduct two extensive human evaluations. Our results confirm the conversational plausibility of the generated dialogs and reveal that people tend to exchange their personal opinions while searching for a product.

In future work, we envision customer-oriented assistant for buying products that assist customers in discussing the merits of products with a sales assistant, grounded in real-world reviews.

%% file: SIGdial23-opinionConv-part7-limitations.tex
\section{Limitations}

As mentioned in Section~\ref{data source and preprocessing}, we focused on the \emph{Cell Phones and Accessories} category of products. However, there is no inherent limitation of our design that prevents future work from including conversations related to other product categories.

Furthermore, an opinion is an observation or a belief that does not need to have evidence to support itself, whereas an argument requires premises. As we discussed in the Section~\ref{participants comments}, study participants expected to have stronger arguments in the generated conversation, rather than only expressing opinions. Therefore, future work should address this aspect utilizing argument mining techniques for generating argumentative dialogues.

%% file: SIGdial23-opinionConv-part8-acknowledgements.tex
\section*{Acknowledgements}
\label{acknowledgements}

This work has been partially supported by the German Federal Ministry of Education and Research (BMBF) within the Junior AI Scientists program under the reference 01-S20060. We would like to thank the anonymous reviewers for their valuable feedback.

%% file: SIGdial23-opinionConv-part9-ethics.tex
\section*{Ethics Statement}
\label{ethics}

Systems designed to influence humans via communication constitute a highly sensitive topic due to their intrinsically social nature \cite{stock2016ethical}. Any automated sales assistant comes along with the ethical risk of not only influencing customer opinion but doing so in ways undesired by customers, e.g., to their financial or otherwise personal disadvantage. Naturally, it is the company that deploys a manipulative sales assistance technology that is at fault, but the question of why research that may be misused in this direction has been undertaken in the first place is still pertinent.

Negotiation differs from persuasion in its goal. Negotiation strives to reach an agreement from both sides, while persuasion merely aims to change one specific person’s attitude and decision \cite{wang2019persuasion}. Most human sales assistants have no interest in deceiving customers, since that very customer may come back to complain, or not come back to buy further products. Modern marketing strategies typically involve building a trustworthy customer relationship which includes the post-purchase stage of the aforementioned customer decision process, where customer satisfaction is to be maximized. We intend our research to serve as a step towards studying the capabilities of automated sales assistance with the goal of mutually beneficial negotiation. Nevertheless, if it turns out that it is easier for technology to manipulate its users with respect to a purchase decision than to consult them for mutual benefit, this must be found out, and publicly, or else no policies against such exploits can be enforced.

Moreover, an automatic sales assistant deployed by a marketplace must be considered separately from, e.g., an automatic sales assistant deployed by an independent third party (including open source variants). We imagine that not only the former will become available in the future, but also the latter, which will be more trustworthy overall.